\documentclass[twocolumn,amsmath,superscriptaddress]{revtex4} 

\usepackage[english]{babel}
\usepackage{graphicx}
\usepackage{enumerate}
\usepackage{subfigure}

\newcommand{\ie}{i.e.,\ }

\newcommand{\pn}{p_\mathrm{new}}
\newcommand{\po}{p_\mathrm{old}}
\newcommand{\rmd}{\mathrm{d}} 
\newcommand{\rme}{\mathrm{e}} 

\begin{document}

\title{
Influence of homology and node-age\\
on the growth of protein-protein interaction networks
}

\author{Arianna Bottinelli} 
\affiliation{Department of Mathematics, Uppsala University, Uppsala, Sweden}

\author{Bruno Bassetti} 
\affiliation{Universit\`a degli Studi di
  Milano, Dip.\ Fisica, Via Celoria 16, 20133 Milano, Italy}
\affiliation{I.N.F.N. Milano, Via Celoria 16, 20133 Milano, Italy}

\author{Marco \surname{Cosentino Lagomarsino}} 
\affiliation{Genomic Physics
  Group, UMR 7238 CNRS ``Microorganism Genomics''}
\affiliation{University Pierre et Marie Curie, 15 rue de l'\'{E}cole
  de M\'{e}decine Paris, France}
\affiliation{Dipartimento di Fisica,
  Universit\`a di Torino, via P.~Giuria, 1, Torino, Italy }

\author{Marco Gherardi} 
\email{marco.gherardi@mi.infn.it}
\affiliation{Universit\`a degli Studi di
  Milano, Dip.\ Fisica, Via Celoria 16, 20133 Milano, Italy}
\affiliation{I.N.F.N. Milano, Via Celoria 16, 20133 Milano, Italy}

\begin{abstract}
  Proteins participating in a protein-protein interaction
    network can be grouped into homology classes following their
    common ancestry. 
    Proteins added to the network correspond to genes added to the
    classes, so that the dynamics of the two objects are intrinsically
    linked.
    Here, we first introduce a statistical model describing the joint
    growth of the network and the partitioning of nodes into classes,
    which is studied through a combined mean-field and simulation
    approach.  We then employ this unified framework to address the
    specific issue of the age dependence of protein interactions,
    through the definition of three different node wiring/divergence
    schemes.  Comparison with empirical data indicates that an
    age-dependent divergence move is necessary in order to reproduce
    the basic topological observables together with the age
    correlation between interacting nodes visible in empirical data.
    We also discuss the possibility of nontrivial joint
    partition/topology observables.
\end{abstract}

\maketitle

\section{Introduction}

The protein-protein interaction (PPI) network represents the physical
interactions between proteins in a cell~\cite{Bork2004}. The
topological properties of this complex network provide an effective
overview of the protein-protein interactions coded by a genome, with
implications for the analysis of signaling and metabolic
pathways~\cite{Maslov2002a}.

In the course of evolution, a genome acquires new genes, and thus new
proteins, by different evolutionary
processes~\cite{Sun2011,Koonin2002}, which include gene duplication
and horizontal transfers. These processes define groups of proteins
with the same common ancestor, termed homology classes. Notably,
homology classes follow well-defined quantitative laws with specific
mathematical
properties~\cite{koonin_plos2011,qian_gerstein,Koonin2002}, dependent
only on genome size and not on further details of a genome's
evolutionary history
\cite{cosentinolagomarsino_sellerio_etal_2009,angelini_amato_etal}.

Following gene duplications~\cite{zhang_2003}, proteins belonging to
the same homology class can modify their binding interfaces to
conserve ancient interactions, lose them, or evolve new ones. This
process generates new PPI network configurations which are subject to
selective pressures of different
kinds~\cite{Heo2011,Wang2011,Teichmann2002}, and allow to construct
increasingly complex biomolecular
machinery~\cite{Pereira-Leal2007,Venkatakrishnan2010,Pereira-Leal2005}.
This mechanism of ``duplication-divergence'' has inspired a thread of
graph-growth modeling work within the physics and computational
biology
community~\cite{R.2002,Pastor-Satorras2003,Vazquez2003,ispolatov_krapivsky,Stein2011,evlampiev_isambert,evlampiev_isambert_2007}.
Generally speaking, these models generate random graph ensembles by
iteratively adding new nodes that are initially copies of existing
ones (and thus interact with all their binding partners) and
subsequently lose and/or rewire interactions by a set of simplified
prescription rules.  This basic mechanism produces graph topologies
resembling empirical PPI networks in many aspects.  Comparison of
model predictions and empirical data leads to the hypothesis that
duplication-divergence can (at least in part) explain PPI network
topologies~\cite{Ispolatov2005,evlampiev_isambert,Presser2008},
starting from the basic observation that duplicate proteins are often
involved in similar protein-protein
interactions~\cite{Pereira-Leal2005,Pereira-Leal2007}.

While it appears that gene duplication plays a role in shaping PPI
networks through evolutionary time~\cite{Navlakha2011}, many questions
remain open. For example, it has been pointed out that the
duplication-age profiles naturally emerging from
duplication-divergence models do not resemble empirical data, and
that, quite reasonably, the availability of binding interfaces could
impose additional relevant
constraints~\cite{kim_marcotte,Sun2012,Talavera2012}. Accordingly,
alternative models have been proposed, where the wiring rules account
for these constraints~\cite{kim_marcotte}.  
Additionally, according to most of these models, ``collapsing''
  multiple homologous neighbors of a protein into one neighbor should
  make the broad degree distribution considerably narrower, which
  does not seem to be the case in empirical data~\cite{berg_lassig_wagner}.
Thus, the actual growth mechanisms of PPI networks is
still under debate and it is unclear how much duplication-divergence
versus other constraints can account for the topology of empirical PPI
networks~\cite{kim_marcotte,Robertson2009,Navlakha2011}.
Additionally, duplication-divergence models typically neglect the
process of homology classes expanding and being formed within a
genome, and thus cannot describe how PPI network links are distributed
among homology classes.  However, the subdivision of genes into
homology classes could constitute another relevant constraint for the
PPI network's structure and should not be neglected a priori.

This work addresses the above issues through a modeling approach. We
consider a (null) statistical graph-growth model describing the joint
growth of PPI network and homology classes structure. The output of
the model is a growing graph, whose nodes are partitioned into
equivalence classes following the empirical size distributions of
protein classes.  The model defines a framework for testing
alternative mechanisms of network growth, where duplication-divergence
can have different weight during the process and thus different
consequences on the final properties of the network.  Within this
setting, we ask about the ingredients that can account for the joint
growth of homology classes and network, as well as reproducing the
main empirical observables such as degree distribution, degree
correlation and correlation between interacting duplication-age
groups.  In our analysis we find in particular that reproducing the
empirical age-correlation between interacting nodes requires a heavy
bias on the duplication-divergence process, which must
  correspond to additional constraints of functional or of physical
  origin.

\section{Background.  \label{section:background}}

\subsection{Network growth by duplication-divergence}

Perhaps the simplest PPI network growth model incorporating the basic
moves of duplication and divergence (DD) was introduced and studied
in~\cite{ispolatov_krapivsky}.  In this model, the network grows by
node duplication and subsequent deletion of some of the duplicate
links with a prescribed probability (divergence).  More precisely, at
each step a randomly-chosen network node is copied, initially
inheriting all the interactions of the original node, and in a second
substep the new node's links are deleted independently with
probability $1-\sigma$.  If no link is left after divergence, the
duplicate node itself is deleted, so that the network remains
connected throughout its evolution.  This process is completely
asymmetric, meaning that the parent node (the one chosen for
duplication) does not lose any connection, and the divergence process
only affects the daughter.  More general variants have been proposed,
for instance by relaxing the requirement of complete asymmetry and of
single-gene duplication~\cite{evlampiev_isambert}, or by introducing
rewiring between existing nodes (which can even become dominant
  in shaping the network~\cite{berg_lassig_wagner}).  For simplicity,
we will restrict to the one-parameter model in the following.

One of the main features of this model is that the described mechanism
leads to an effective preferential attachment principle, since
high-degree nodes are more likely to have a neighbour being duplicated
by random choice.  Specifically, the probability of a new link being
attached to a node of degree $k$ is proportional to $k$.  As a
consequence, the degree distribution of the growing network develops
power-law tails $\sim k^{-\gamma}$ for large
degrees~\cite{ispolatov_krapivsky}. Exponents in the range
$\gamma\in[2,3]$ are realized by choices of $\sigma\in(0,1/2]$.
Comparison with available subsets of empirical PPI networks yields
values of the link-retention probability $\sigma$ around
$0.40(\pm0.05)$ for \emph{S.\ cerevisiae}, \emph{D.\ melanogaster} and
\emph{H.\ sapiens} \cite{ispolatov_krapivsky}.  
The average total number of links $L(N)$ as a
function of the network size $N$ can also be predicted by mean-field
calculations (see Section~\ref{section:meanfield}).

\subsection{Homology class partitioning by the Chinese restaurant process}
Duplication plays a fundamental role in the evolution of homology
classes as well~\cite{cosentinolagomarsino_sellerio_etal_2009}, as it
constitutes the main drive for class expansion, at least in
eukaryotes.  Equally, a genome ``innovation'' move (for instance by
horizontal transfer) causes the creation of new homology classes.

A simple class of partitioning processes incorporating the basic moves
of class expansion and innovation is capable of explaining the scaling
laws observed in domain-class partitioning~\cite{angelini_amato_etal}.
The paradigm of these models is the so-called ``Chinese Restaurant
Process'' (CRP)~\cite{bassetti_zarei_etal,DS05, pitman_book2006,qian_gerstein},
which is the one that will be used here.
In this process, at each iteration the genome goes from having $n$ to $n+1$
genes, and either a new class is created (with probability $\pn$) or a
domain is added to an existing class (with probability $\po=1-\pn$).
A crucial ingredient of the CRP is the dependence of $\pn$ and $\po$
on the size of the growing proteome, whose effect is to reproduce in
the model the observed sublinear scaling of the number of domain
classes $F(N)$ with genome size $N$:
\begin{equation}
\label{eq:def_probcrp}
\begin{aligned}
\pn &= \frac{\alpha F(N)+\theta}{N+\theta},\\
\po &= \frac{N-\alpha F(N)}{N+\theta},
\end{aligned}
\end{equation}
where $\alpha\in(0,1)$ and $\theta\geq 0$ are parameters of the model.
(The extreme cases $\alpha=0,1$ could be included, but we will neglect
them here for clarity.)
The per-class probability of duplication
is defined as
\begin{equation}
\label{eq:perclassprobcrp}
\po^{(i)}=\frac{j_i-\alpha}{N+\theta},
\end{equation}
where $j_i$ is the size of the $i$-th class. This corresponds to an
asymptotically uniform extraction, which realizes an effective
preferential attachment principle.  The parameter $\alpha$ describes
the dominance of innovation over duplication, while $\theta$ is a
fixed size scale at which preferential attachment sets in.  Mean-field
calculations, supported by simulations, show
\cite{angelini_amato_etal} that the asymptotic behaviors of the
class-size distribution $f(j,N)$ and of the total number of classes
$F(N)$ are
\begin{equation}
\label{eq:crp_asymptotic}
\begin{aligned}
  f(j,N) &\sim j^{-(1+\alpha)},\\
  F(N) &\sim N^\alpha,
\end{aligned}
\end{equation}
for large $N$ and $j$.  As a consequence, $\pn$ and $\po$ scale
as
\begin{equation}
\label{eq:asymptotic_probcrp}
\begin{aligned}
\pn&\sim\alpha N^{\alpha-1},\\
\po&\sim 1-\alpha N^{\alpha-1}.
\end{aligned}
\end{equation}
These predictions are in good qualitative agreement with empirical
data for prokaryotic
proteomes~\cite{angelini_amato_etal,cosentinolagomarsino_sellerio_etal_2009}.

\section{Model and Methods}

\subsection{\label{section:model_definition}Definition of a
  statistical model combining genome partitioning and network growth}

As we discussed, from a simplifying perspective, the growth of PPI
networks and genome partitioning in homology classes are produced by
essentially the same basic evolutionary moves of innovation and
duplication on the genes.  For this reason, the model proposed here is
defined by abstract realizations of these basic moves on the level of
both the network and the homology classes.  This is achieved by a
simple coupling between the duplication-divergence model of network
growth and the CRP partitioning, as reviewed in the Background
section.  In particular, a class expansion move is associated with a
network duplication move, and a proteome innovation move with a
network move wiring the new node to the existing network.  Thus, the
model could be termed ``Duplication Divergence Innovation Wiring''
(DDIW), and describes the growth of homology classes and PPI network
jointly.

\begin{figure}
\centering
\includegraphics[scale=.8]{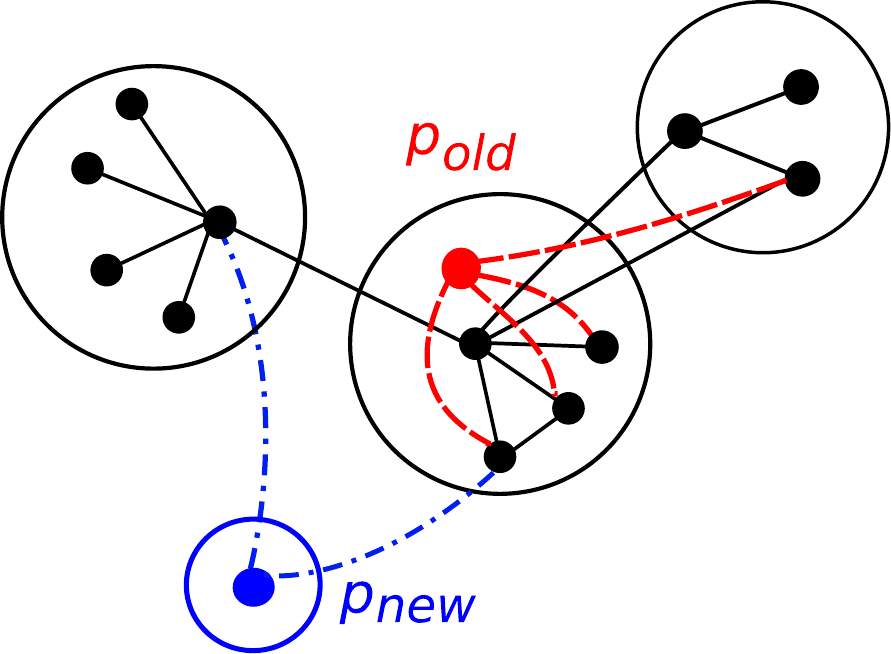}
\caption{(Color online) Illustration of the moves in the DDIW model.
At each step either a new class containing one node is added,
and the new node is linked to one or more existing nodes (innovation-wiring,
done with probability $\pn$),
or a randomly-chosen node is duplicated inside a class and
the replica's links activated independently with probability $\sigma$ (duplication-divergence,
done with probability $\po$).
Filled circles are nodes, lines are links;
large circles are homology classes;
the red node and its dashed links are the results
of a duplication-divergence move;
the blue node and its dot-dashed links are the results
of an innovation-wiring move.
\label{figure:model}}
\end{figure}

Let $\pn$, $\po^{(i)}$, and $\po=\sum_i \po^{(i)}$ be defined as in
(\ref{eq:def_probcrp}) and (\ref{eq:perclassprobcrp}) in terms of the
number of classes $F(N)$ and the size of the $i$-th class
$j_i$.  The basic data structure of the model includes the
topology of the PPI network, and the information on the partitioning
of its nodes (see Figure~\ref{figure:model}).  Given a
proteome/network of size $N$, the growth process is defined by the
following two rules acting on the classes and on the graph topology.
\begin{itemize}
\item[1.]{}  {\bf a:~DUPLICATION (classes)} Choose a class $i$ with
  probability $\po^{(i)}$, and duplicate a randomly-chosen target node
  inside class $i$.\\
  {\bf b:~DIVERGENCE (network)} Attach the new node
  to each of the target's neighbors independently with probability
  $\sigma$.
\item[2.]{} {\bf a:~INNOVATION (classes)} Otherwise (\ie with
  probability $\pn$), create a new node in a new class.\\
  {\bf b:~WIRING (network)} Attach the new node to one or more nodes
  in the existing network, independently of their classes. (The
    additional rules describing this step are listed in Sec.~\ref{section:modelvariants}.)
\end{itemize}

Altogether, there are three parameters governing the dynamics,
$\alpha\in(0,1)$, $\theta\geq0$, and $\sigma\in(0,1]$.  Notice that,
while the network dynamics is dependent on the configuration of the
partitioning, the evolution of the latter is not affected by what
happens at the network level.  Therefore, partitioning is assured by
definition to reproduce the CRP predictions for all choices of the
parameters.  Notice that class-expansion can also occur by horizontal
transfer of members of an existing homology
class~\cite{rocha_treangen}, but we will disregard this process
here. In fact, while this mechanism is widespread in bacteria, we
found that there was no need to incorporate it explicitly in the model
in order to have a good fit with data for both networks and homology
classes.

Technically, we choose a slightly different divergence rule from the
model of ref.~\cite{ispolatov_krapivsky}.  In order for duplication to
always be successful (\ie no node being left without any links) we
impose a randomly chosen link to be conserved, and divergence to be
performed on the remaining ones, i.e. the model assumes that each
duplicated node is preserved by selection and cannot be disconnected
from the existing network.  The same hypothesis holds for the original
model, but is implemented by removing the disconnected nodes.  The
different implementation implies that the divergence rule explained in
Sec.~\ref{section:background} yields a degree-dependent probability of
duplication, since less connected nodes are more prone to have all
their links disconnected; the rule used here, instead, assigns the
same probability of duplication to every node.  Despite this bias, the
modified model incorporates the same basic mechanisms as the previous
one, and we verified that it leads to the same qualitative results
(some features match also quantitatively, see
Sec.~\ref{section:meanfield}).  The main rationale behind this choice
is a simplification of the mean-field equations, as it makes it
unnecessary to estimate the number of deleted nodes.

The initial condition will be chosen as the complete $3$-graph,
  which is the smallest non-bipartite network.  Results do not change
  appreciably by starting with different small networks (we did not
  study systematically the dependency of the results from initial
  conditions built as large networks).
  
 We choose to exclude
  self-interactions from the model, as they play a biologically
  distinct role in the network, and they probably deserve to be
  considered separately~\cite{Venkatakrishnan2010}.

\subsection{\label{section:modelvariants}Model variants allowing to study the effect of different
  growth mechanisms on the topology}

The wiring rule is not completely specified by the definitions above.
Its implementation will be given in the following.
At the network level, the rules concerning the topology can be modified without affecting
the basic structure of the model. Here, we study a minimal version
and consider different variants for such rules, which allow to address the recently formulated problem of
the age-dependency of empirical interactions
~\cite{kim_marcotte}.

We start focusing on the wiring move.
Once introduced, the new node can be attached to a single node chosen in the existing network by a
preferential attachment (PA) or anti-preferential attachment (AP)
principle with respect to the old node's degree. 
The former alternative describes the tendency of new, specialized 
proteins to interact more likely with old proteins that perform basic tasks,
 the latter reflects the relationship between the binding probability 
 and the available interaction surface of existing
nodes~\cite{kim_marcotte}.
Alternatively, the new node can be wired to a size-dependent or
configuration-dependent number $l$ of existing nodes.

\begin{figure}
\centering
\includegraphics[scale=0.57]{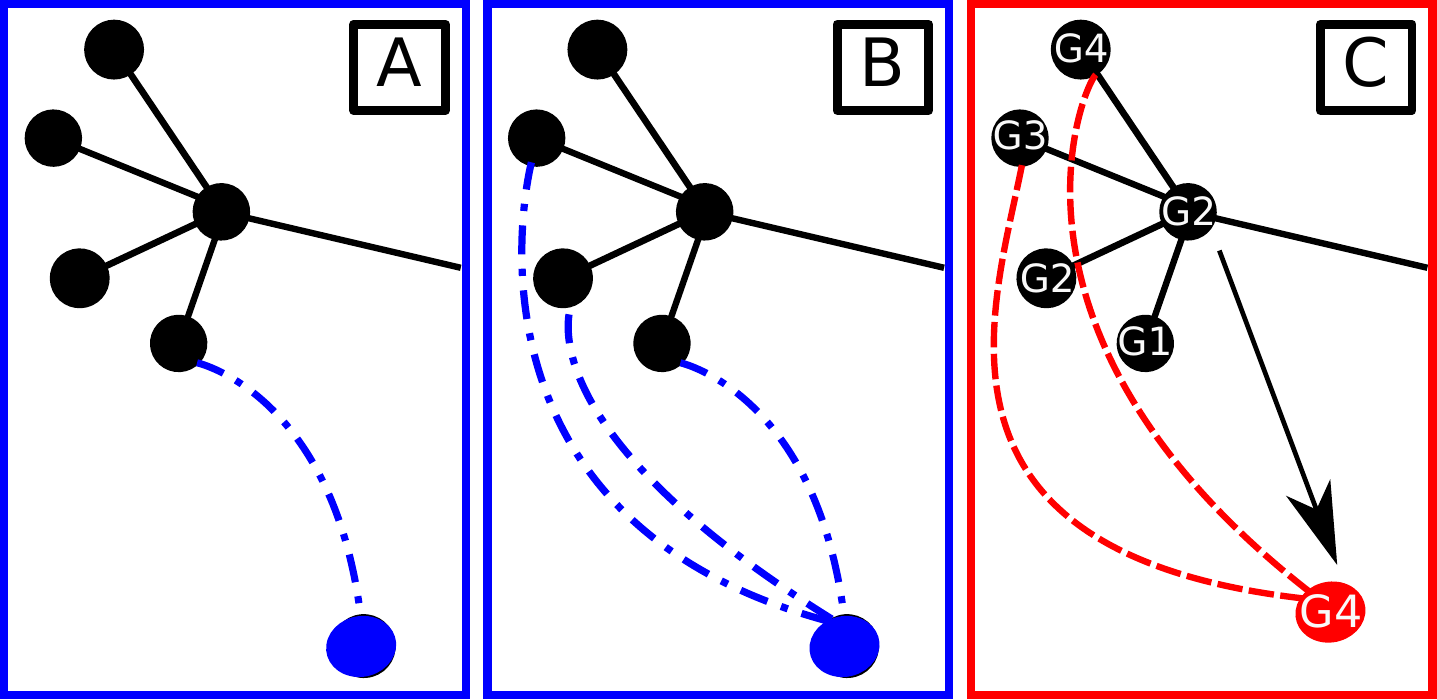}
\caption{(Color online) Variants of the model.
Symbols and colors have the same meaning as in Fig.~\ref{figure:model}.
(A) DDIW+AP (anti-preferential-attachment innovation with a single link).
During innovation, the new node carries one new link, whose target
node is chosen with probability inversely proportional to its degree.
(B) DDIW+EAP (anti-preferential-attachment innovation, with multiple links).
During innovation, the new node carries a number of links
proportional to the current average degree.
(C) A-DDIW (age-dependent divergence).
During divergence of a duplicated node, the probability of keeping a link depends on
the difference in age between the two nodes linked
(higher age differences corresponding to lower probabilities).
\label{figure:model_variants}
}
\end{figure}

Other modifications are possible for the divergence move, for example
by making the link-retention probability $\sigma$ depend on the current
configuration of the network or on the age difference between the two nodes that are connected by the link considered by divergence.  
Here we consider three main variants
\footnote{We have also studied the variant DDIW + PA, where the new node is wired to the old network according to a preferential attachment principle, but the results did not show a significant difference from the DDIW + AP variant described here, and thus will not be reported.} (see Fig.~\ref{figure:model_variants})
\begin{enumerate}[A]
\item\label{variant_AP} DDIW + AP. The wiring move establishes a
  single new link between the new node and an existing node $i$ of
  degree $k_i$, chosen with probability proportional to $1/k_i$.  This
  anti-preferential rule reflects the growing of the binding
  probability with the interaction surface available.
\item\label{variant_nAP} DDIW + extensive AP (EAP). The wiring
  move attaches the new node to $l=\left[\gamma\left<k\right>\right]$
  existing nodes, chosen with anti-preferential attachment;
  $\left[\gamma\left<k\right>\right]$ is the closest integer to a
  fraction $\gamma\in(0,1)$ of the mean degree in the present
  configuration.
\item\label{variant_age} Age-dependent DDIW (A-DDIW). The wiring move
  is the same as in \ref{variant_AP}.  The divergence step implements
  a kind of preferential attachment which takes into account the
  node's age in the following way.  Let $a_i$ be the age of node $i$,
  \ie the number of iterations the process underwent since the node
  was born.  A link to node $i$ inherited from the target node is kept
  with probability $1$ if $a_i<\sigma N$, where $N$ is the size of the
  network, and with probability $0$ otherwise.  This rule implements
  non-neutral selective pressure towards maintaining ancient
  well-established basic cellular machinery.
\end{enumerate}

\subsection{\label{section:data_analysis}Empirical Data Sets and Data
  Analysis Methods}

Data for protein binding is obtained from the most recent (october
2011) \emph{Database of Interacting Proteins} (DIP) \cite{DIP}.  We
filter out self-interactions between single proteins and interactions
between proteins expressed by different genomes; different strains are
considered as different organisms.  Moreover, we exclude all virus
data, and all networks with less than $10$ nodes.  We end up with $1$
\emph{archaeon}, $14$ \emph{bacteria}, and $7$ \emph{eukaryotes}; a
list of all organisms considered in the study of network topology is
presented in Table~\ref{table:genomes}, together with the observed
number of proteins $N$ and interactions $L$.  Notice that the networks
we can construct from DIP only include subsets of the full proteomes.
For example the \emph{C.~elegans} network in our dataset is smaller
than that of \emph{S.~cerveisiae}, despite its genome being much
larger, possibly creating significant under-sampling problems in
  the data.  See Sec.~\ref{section:discussion} for a discussion of
this issue.

Homology classes are built starting from the SUPERFAMILY database for
domain assignment \cite{SUPERFAMILY}.  We reconstruct the domain
architectures as ordered lists of domains and gaps; a gap is defined
as a subsequence of 100 or more ``AA'' not scored for domain
\cite{Fusco_et_al_2008}.  Two proteins are in the same homology class
if their architectures are exactly matching.  We also tested a more
relaxed criterion (allowing for repetitions of domain architectures),
and obtained the same results as those presented in the following for
the stricter criterion.
Moreover, we also considered data restricted to longest transcripts
in eukaryotes, finding no difference in the scaling
(we remark that longest-transcript data in the dataset are very
incomplete, so we will not include them in the forthcoming analysis).
We filter out genomes with more than $19\,000$ assignments;
altogether, we work with data for $1384$
organisms --- $87$ \emph{archaea}, $1077$ \emph{bacteria}, and $220$ \emph{eukaryotes} 
--- for the homology classes, but only $22$
networks with sufficiently large sampling of the interactions.

Beside network topology and homology classes, we are interested in
evolutionary ages of proteins. 
For the proteome of \emph{S.\ cerevisae}, we
use data from Wapinski \emph{et al.\
}\cite{wapinski_pfeffer_etal_2007}, where duplication events for a
number of genes of \emph{S.\ cerevisiae} are divided into ten classes,
labeled A, B, C, D, E, WGD, G, H, I, J, depending on when in the
evolutionary history of \emph{Ascomycota} they occurred (class A being
the more recent). We further group these classes into four
superclasses (labeled G1--4), keeping the whole-genome duplication
(WGD) alone, due to the abundance of its elements:
\begin{equation*}
\begin{aligned}
G1 &= I + J\\
G2 &= G + H\\
G3 &= WGD\\
G4 &= A+B+C+D+E.
\end{aligned}
\end{equation*}
By this procedure, we assign $210$ genes to age group $G1$, $85$ to
$G2$, $691$ to $G3$ (WGD), and $91$ to $G4$.  The age of a protein is
defined as the superclass of the oldest duplication event in which it
is reported to be involved.  It should be noted that the WGD has a
different phenomenology than the single-gene duplication events
considered here;  we do not exclude it from our data, but its
modelization is out of the scope of the present work
(see \cite{evlampiev_isambert_2007}).
In order to
evaluate the history dependency of protein interactions, we use the
interaction density $D_{m,n}$ between two age groups $m$ and $n$ as an
indicator of age correlation. It is defined, following
\cite{kim_marcotte}, as
\begin{equation}\label{eq:interactiondensity}
D_{m,n}=\log_2\left[\frac{L_{m,n}}{E_{m,n}}\frac{N(N-1)}{2L}\right],
\end{equation}
where $L_{m,n}$ is the number of links between the age groups $m$ and
$n$ and $E_{m,n}$ is the number of possible links between nodes of the
two groups, which only depends on the number of nodes in $m$ and $n$.
The average interaction density gradient, defined as \cite{kim_marcotte}
\begin{equation}
\Delta D = \sum_{n=2}^4\sum_{m<n}\left(D_{m+1,n}-D_{m,n}\right),
\end{equation}
measures the overall correlation present between the ages of proteins;
a positive value indicates that newer nodes preferentially link
with newer nodes. We will use the sign of $\Delta D$ as a marker
of correlation or anti-correlation between ages.

Fits of data against non-linear analytic expressions are performed
by minimization of the squared residuals through the 
standard Levenberg-Marquardt method, and are systematically
checked for stableness under the introduction of a cutoff on
small-size data.

\section{Results}  

We ask in which conditions the model or its variants fulfill the
following requirements.  Firstly, it should qualitatively reproduce the features
of both the duplication-divergence and CRP ``pure'' models. Secondly,
it describes the enriched data-structure of network plus homology
classes, and it should predict the behavior of joint topology-partition
observables, including history-dependency of interactions.

All variants of the DDIW model reproduce the same homology-class
scaling as the pure CRP, essentially because the class partitioning is
not affected by the network dynamics by definition.
A simple scaling argument suggests that the duplication-divergence
predictions are expected to be recovered for large $N$, since the scaling of $\pn$
and $\po$, Eq.~(\ref{eq:asymptotic_probcrp}), shows that
duplication becomes dominant in this regime.  Therefore, the model is
expected to behave as pure duplication-divergence in the large-$N$
limit; it remains to clarify what happens at intermediate values of $N$.
In the following subsections we address some of these questions;
the large-$N$ behavior is clarified by mean-field techniques,
while finite values of $N$ are studied by means of numerical simulations.
The analysis of how the partitioning into homology classes correlates
with the network structure will be briefly addressed to in Sec.~\ref{section:discussion},
but its systematic study will be left to future work.

\subsection{\label{section:meanfield}Mean-field theory accurately
  predicts scaling of the total number of links}

Mean-field calculations give reliable estimates for the behavior of
the duplication-divergence network growth model and for the
class-expansion innovation model
separately~\cite{ispolatov_krapivsky,angelini_amato_etal}, therefore
it makes sense to apply the same procedure to the joint model.  The
mean-field approach essentially consists in neglecting the
fluctuations due to the statistical nature of the models and writing
``macroscopic'' differential equations for the average quantities,
which can be treated analytically.  In this section we will use this
tool to study the average total number of links $L(N)$ as a function
of the number of nodes $N$ for the variants of the joint-evolution model
described in the previous section.  In principle, other
characteristics of the network may be accessible through mean-field
calculations, such as the degree
distribution, but we will not treat them here.

For the duplication-divergence model alone (in the variant defined
in section~\ref{section:model_definition}), 
the simplification we introduced allows to write a
slightly more general expression for $L(N)$ than that obtained in
\cite{ispolatov_krapivsky}.  Let $N_k$ be the average number of nodes
with $k$ links in a network of size $N$ (the average is intended on
all realizations of the stochastic process up to size $N$). Clearly,
\begin{equation}
\label{eq:identity_N}
\sum_k N_k = N,
\end{equation}
and
\begin{equation}
\label{eq:identity_L}
\sum_k kN_k = 2L(N),
\end{equation}
where the sums are extended to all possible values of the degree $k$
(say, from $1$ to $\infty$).  
$L(N)$ varies at each duplication
following the mean-field equation
\begin{equation}
\label{eq:meanfield_equation_DD}
\Delta L(N) \simeq \sum_k \frac{N_k}{N}\left[1+(k-1)\sigma\right],
\end{equation}
where $\Delta L(N)\equiv L(N+1)-L(N)$.  The summand takes into account
the duplication of a node of degree $k$, which is performed with
probability $N_k/N$.  The term in square brackets reflects the fact
that by definition at least one of the links is maintained, while the
other $k-1$ links are kept independently with probability $\sigma$.
Performing the sum by applying identities (\ref{eq:identity_N}) and
(\ref{eq:identity_L}) yields
\begin{equation}
  \label{eq:ODE_DD}
  \Delta L(N) \simeq (1-\sigma)+2\sigma\frac{L(N)}{N}.
\end{equation}
This can be approximated by the following (large $N$) differential
equation
\begin{equation}
  \frac{\rmd L}{\rmd N} \simeq (1-\sigma)+2\sigma\frac{L}{N}.
\end{equation}

\begin{figure}
\centering
\includegraphics[scale=0.35]{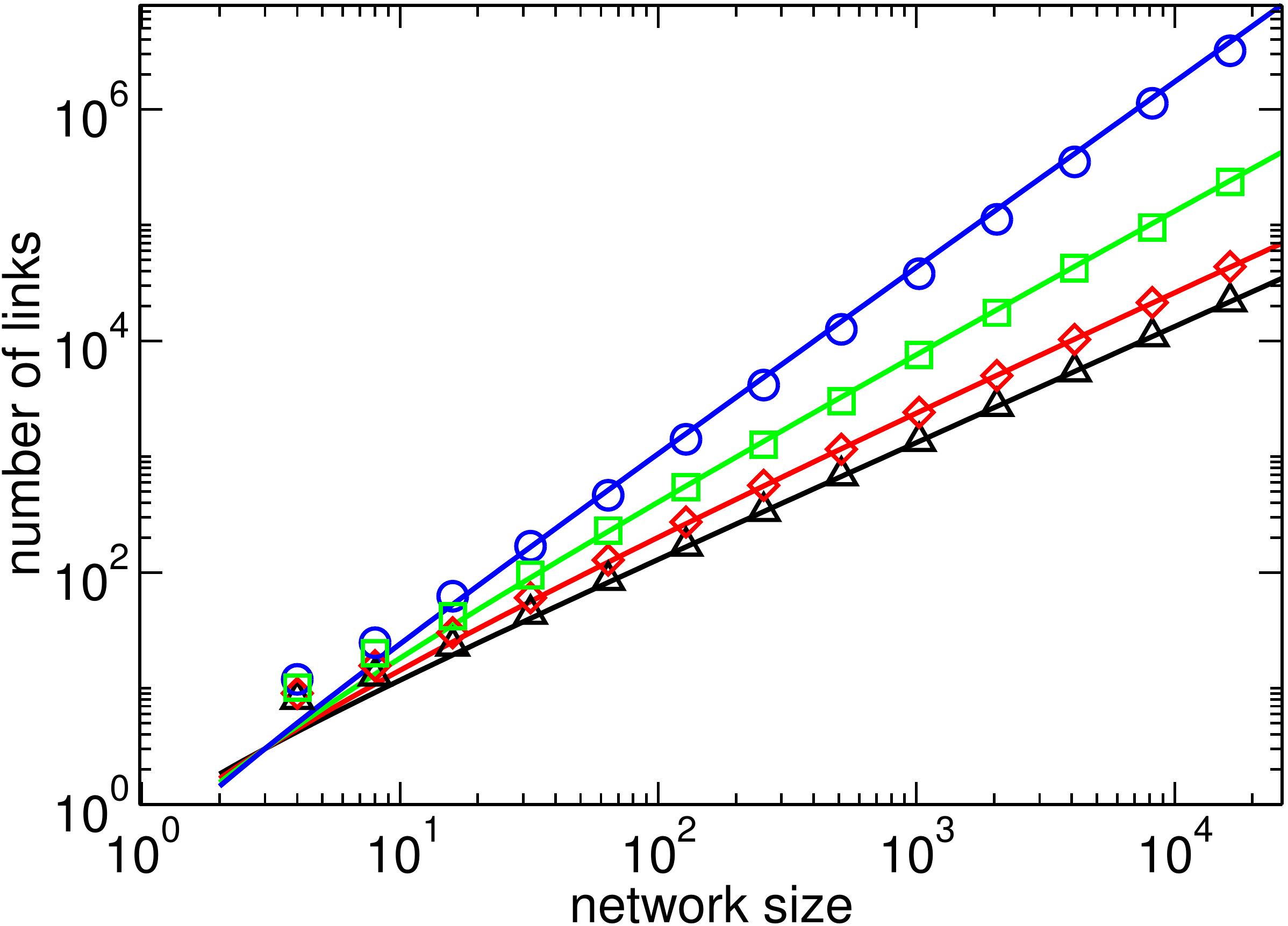}
\caption{ (Color online)
Average total number of links $L(N)$ as a function of
  network size   
for the pure duplication-divergence model.
Solid lines show the mean-field prediction, while symbols are the
results of numerical simulations ($100$ realizations);
error bars are smaller than symbols.
Triangles correspond to $\sigma=0.2$, diamonds to $0.4$, squares to $0.6$, circles to $0.8$.
\label{figure:meanfield_LN_DD}}
\end{figure}

Solving this equation with a formal initial condition $L(N_0)=L_0$
gives the solution
\begin{equation}
  \label{eq:meanfield_solution_DD}
  L(N) \simeq \frac{1-\sigma}{1-2\sigma}N- 
  \left(L_0 - \frac{1-\sigma}{1-2\sigma}N_0 \right)\left(\frac{N}{N_0}\right)^{2\sigma}.
\end{equation}
In the following, we will fix the initial condition to the complete 3-graph ($L(3)=3$),
in order to avoid the proliferation of irrelevant parameters.
The presence of two regimes is apparent, where the first or the second
term dominate, corresponding to $\sigma<1/2$ and $\sigma>1/2$
respectively.  Notice the alternating-sign pattern of the corrections
to scaling, which can cause the observation of a small-size effective
exponent higher than both $1$ and $2\sigma$ (see
Sec.~\ref{section:universal_fits}).  By taking the limit $\sigma\to
1/2$ one has $L(N)=1/2(N\log N) + \mathcal{O}(N)$, thus recovering the
three different regimes of the original DD
model~\cite{ispolatov_krapivsky}.  Figure~\ref{figure:meanfield_LN_DD}
shows that mean-field predictions correctly reproduce the results of
simulations, even for fairly small values of $N$; small deviations from
mean-field appear only for large values of $\sigma$, which are not
very relevant empirically, as the link density would be too high
compared to empirical data.

\begin{figure*}
 \centering
   \subfigure
    {\includegraphics[scale=0.3]{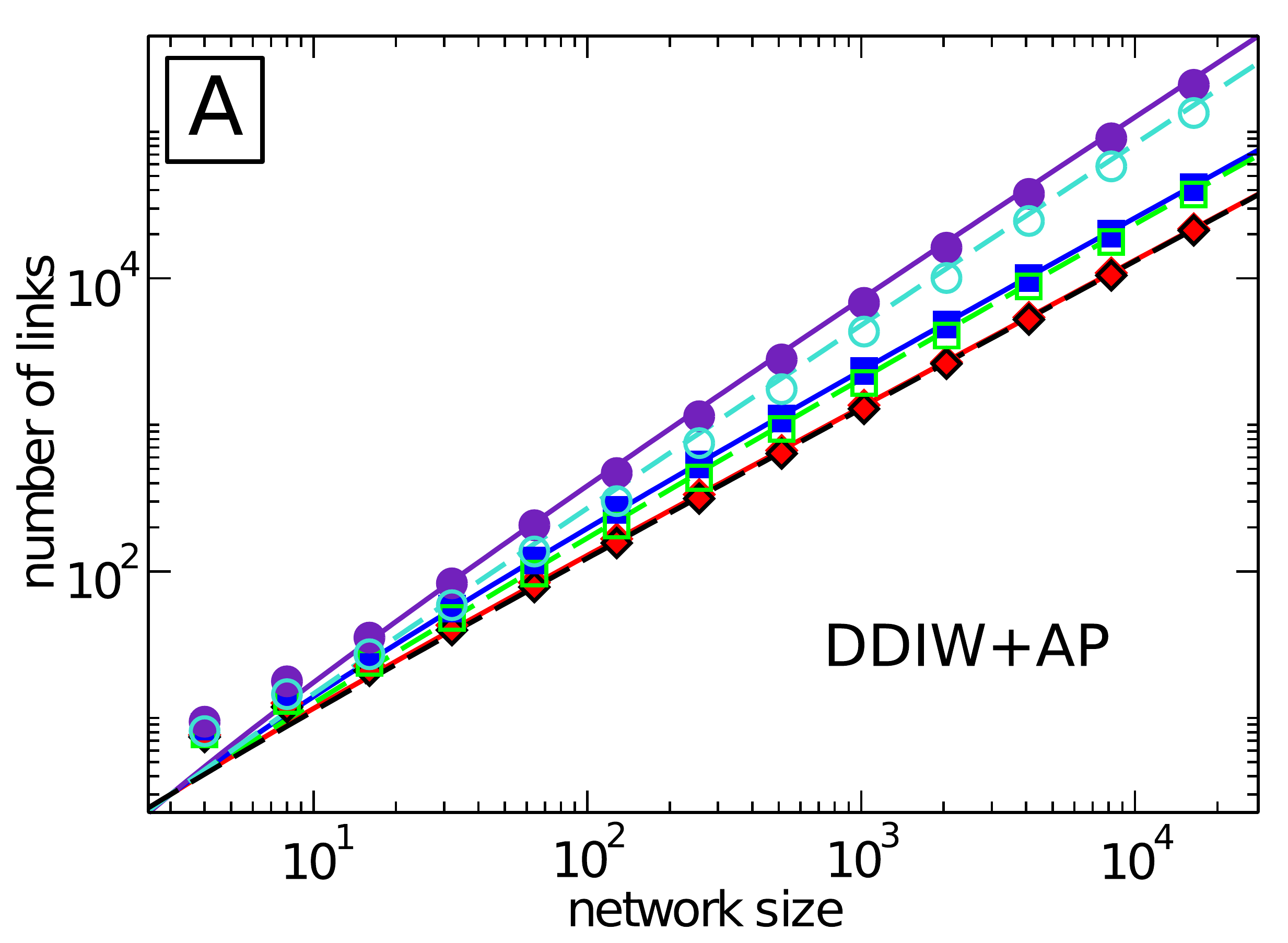}}
     \subfigure
    {\includegraphics[scale=0.3]{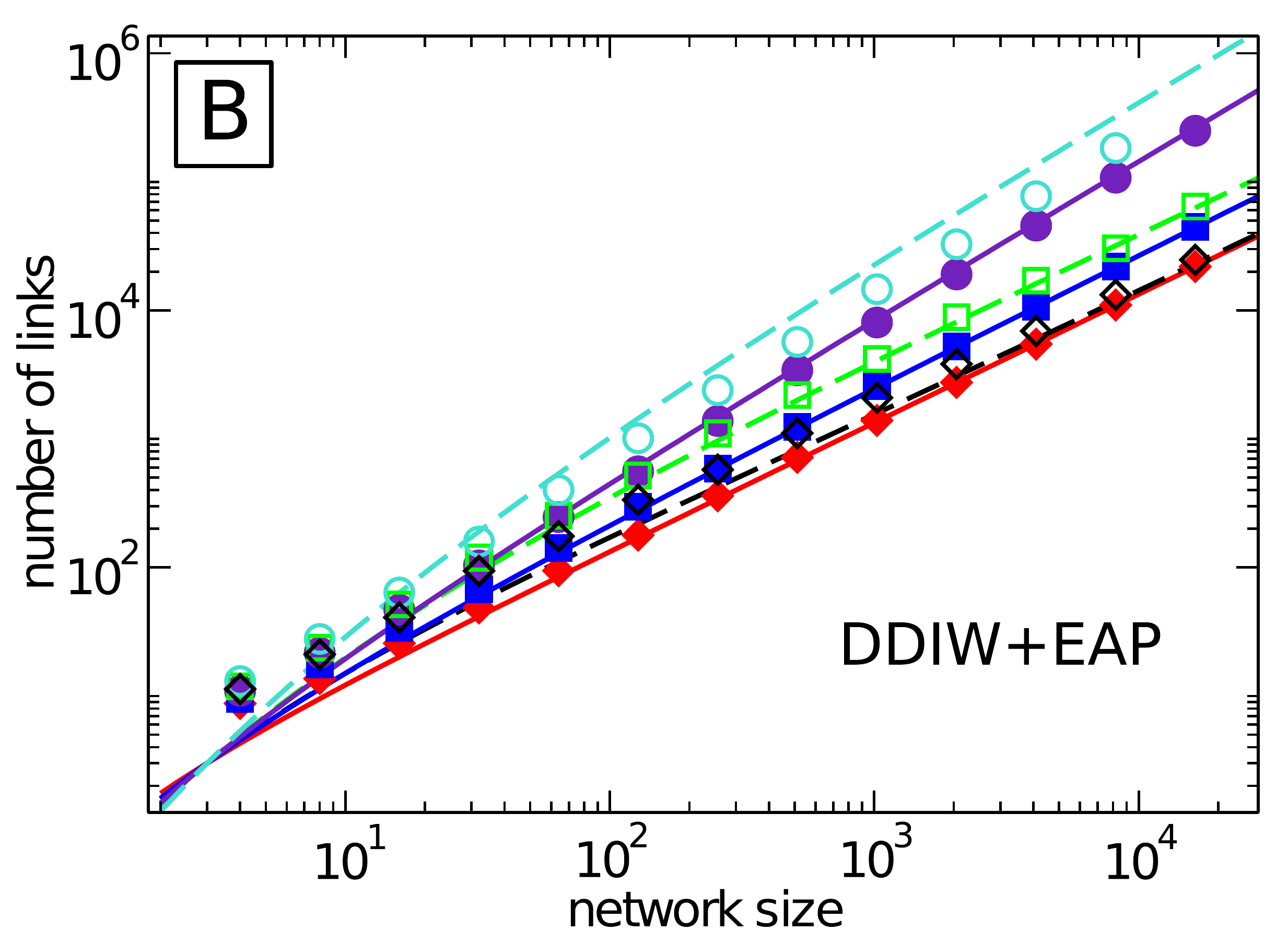}}
    \subfigure
    {\includegraphics[scale=0.3]{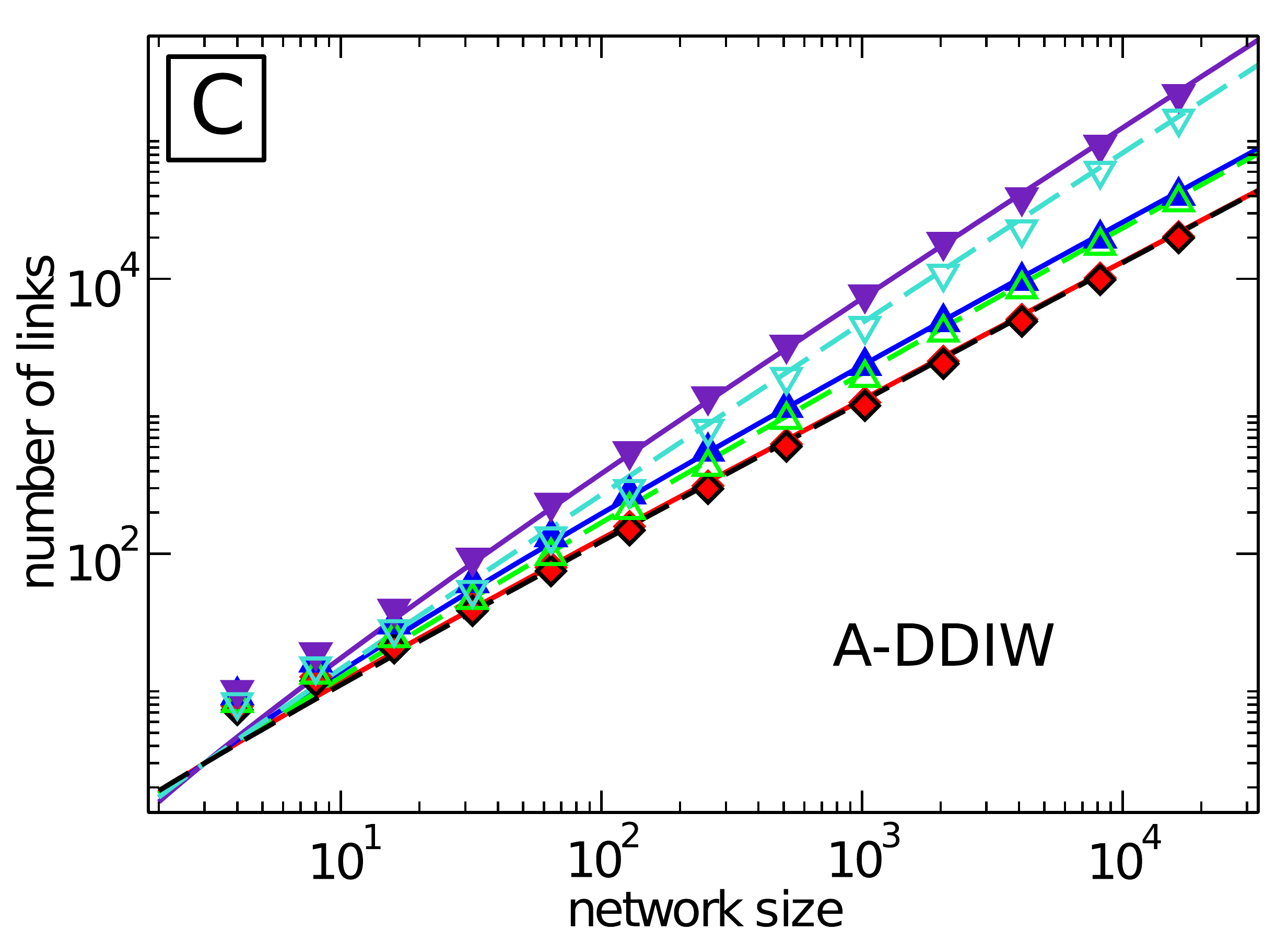}}
    \subfigure
    {\includegraphics[scale=0.218]{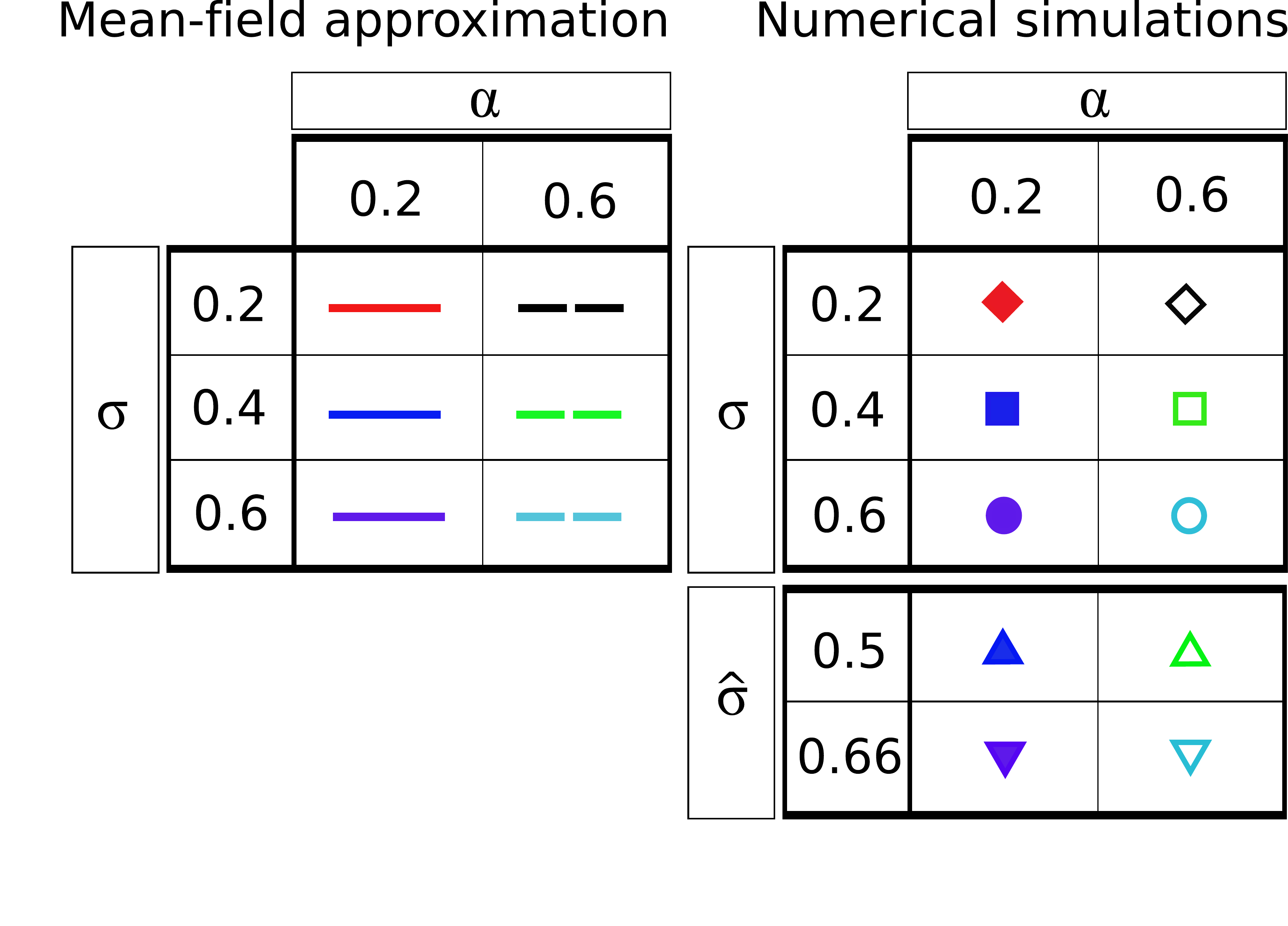}}
\caption{(Color online) 
The scaling of the average total number of links $L(N)$ as a function of network 
size is captured by simple mean-field estimates for all model variants.
Solid curves show the mean-field prediction, while symbols are numerical results
averaged over $100$ realizations.
(A) Variant DDIW+AP (anti-preferential attachment innovation move): the mean field estimate agrees with the simulation results.
(B) Varian DDIW+EAP (anti-preferential attachment innovation move with extensive number of links): deviations are present for the larger values of $\alpha$, but there is good agreement between mean-field estimate and simulations.
(C) A-DDIW (age-dependent duplication divergence): in this case, the mean-field estimates with the same slope of panel (A) are valid for simulations with a rescaled value $\hat\sigma$ of the parameter $\sigma$, related to link retention (note that $\sigma$ is not the link retention probability in this variant, see text).    
\label{figure:meanfield_LN_DDCRP}}
\end{figure*}

We now consider the different variants of the joint DDIW model.  The
increase in the total number of links at each step is given either by
$l$ (if the innovation move is chosen) or by the same sum as in
(\ref{eq:meanfield_equation_DD}), if the duplication move is chosen.
We will not consider variant \ref{variant_age}, since in this case
solving the mean-field equation for the number of links requires
knowledge of the node-age distribution in the network.  Thus, for the
first two variants we have
\begin{equation*}
\Delta L(N)\simeq\pn l(N)+\po\sum_k \frac{N_k}{N}\left[1+(k-1)\sigma\right],
\end{equation*}
where $l(N)$ is the average of $l$ over realizations of the process up
to size $N$.  By plugging in the asymptotic forms
(\ref{eq:asymptotic_probcrp}) and taking the continuum approximation
as in (\ref{eq:ODE_DD}), we obtain
\begin{equation}
\label{eq:ODE_DDCRP}
\begin{aligned}
\frac{\rmd L}{\rmd N} &\simeq \alpha N^{\alpha-1}l(N)\\
&\phantom{\simeq}+\left(1-\alpha N^{\alpha-1}\right)\left(1-\sigma+2\sigma\frac{L}{N}\right),
\end{aligned}
\end{equation}
which has to be solved separately for the two cases $l(N)=1$ (DDIW+AP,
variant \ref{variant_AP}) and $l(N)=\gamma 2L/N$ (DDIW+EAP, variant
\ref{variant_nAP}).  The solution is presented in some detail in
the Appendix, and we concentrate here on the asymptotic behavior. 
Up to exponential corrections of the form $\exp(x^{-\eta})$ with $\eta>0$,
the number of links scales as
\begin{equation}
\label{eq:asymptotic_LN_AP}
L(N)\sim a N^{2\sigma} + b N
\end{equation}
for variant \ref{variant_AP}, and as
\begin{equation}
\label{eq:asymptotic_LN_nAP}
L(N)\sim c N^{2\sigma} + d N^\alpha + e N
\end{equation}
for variant \ref{variant_nAP}; $a$ and $b$ are functions of $\sigma$
and $\alpha$, while $c$, $d$, and $e$ are functions of $\sigma$,
$\alpha$, and $\gamma$.  The exponential corrections are proportional to $\exp(\pn)$, 
which indicates the influence the partitioning
process has on the early stages of the growth process.
Figure~\ref{figure:meanfield_LN_DDCRP} shows a comparison between
mean-field results and numerical simulations.  Deviations are apparent
for $(\sigma,\alpha)=(0.6,0.6)$ and $(0.2,0.6)$
in the DDIW+EAP variant, but
theoretical predictions are accurate for other values and for the
DDIW+AP variant.  The structure of the power-law corrections to
scaling is similar to that of the pure DD model and as long as
$\alpha<2\sigma$ (which is the case for the universal fits to
empirical data presented in Sec.~\ref{section:universal_fits}) the
asymptotic behavior only depends on $\sigma$, up to the sub-leading order.
This suggests that the scaling behavior of the hybrid DDIW model is to
a certain extent robust with respect to the details of the innovation dynamics.

Concerning the scaling of the number of links in variant
\ref{variant_age} (A-DDIW, Fig.~\ref{figure:meanfield_LN_DDCRP}), 
note that in this case the definition of $\sigma$ does
not allow to interpret this parameter  as the average fraction of links
retained after node duplication.  This is due to a non-trivial correlation between 
node age and node degree, which is not straightforward to include in the
mean-field calculation.  
Nevertheless, numerical simulations indicate that the
asymptotic  behaviour of $L(N)$ derived for variant \ref{variant_AP} (DDIW+AP) also holds
for variant \ref{variant_age}, up to a rescaling of $\sigma$. This can
be seen in Fig.~\ref{figure:meanfield_LN_DDCRP}, where the mean-field predictions
are compared with numerical results for the rescaled values  $\hat\sigma$.

\subsection{\label{section:universal_fits}Scaling of the number of
  links and classes as functions of genome size is reproduced by
  universal parameters independent of the model variant}

Having established that the scaling for the number of links is
captured by simple mean-field estimates and indicates well-defined
parameter regimes, we constrain the parameters by
comparing to the available empirical data.
Specifically, we fix the three parameters $\alpha$, $\theta$, and
$\sigma$ by fitting the mean-field expressions against data for
homology classes and for PPI networks.

\begin{figure}
  \centering
  \subfigure
    {\includegraphics[scale=0.3]{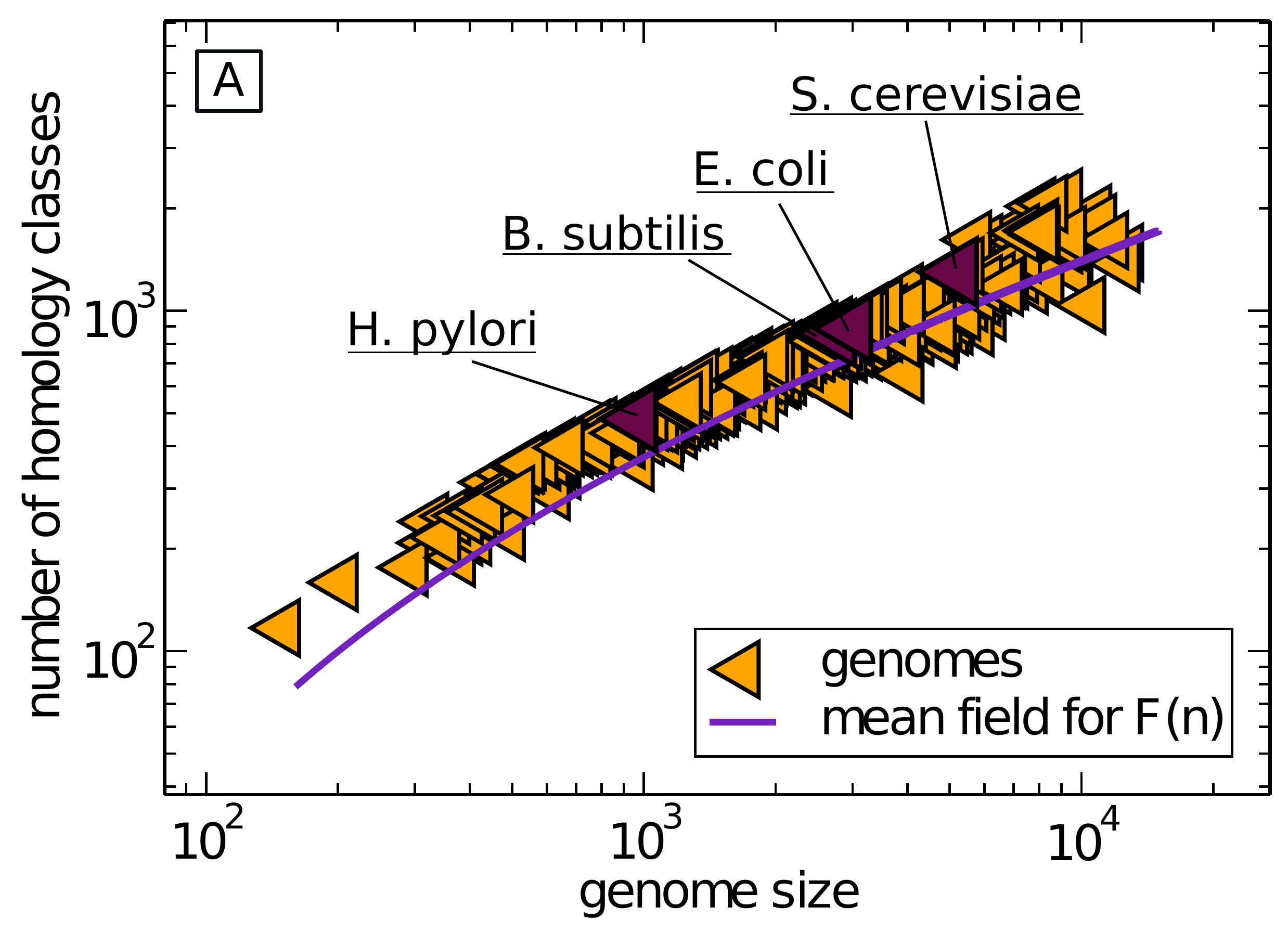}}
  \subfigure
    {\includegraphics[scale=0.3]{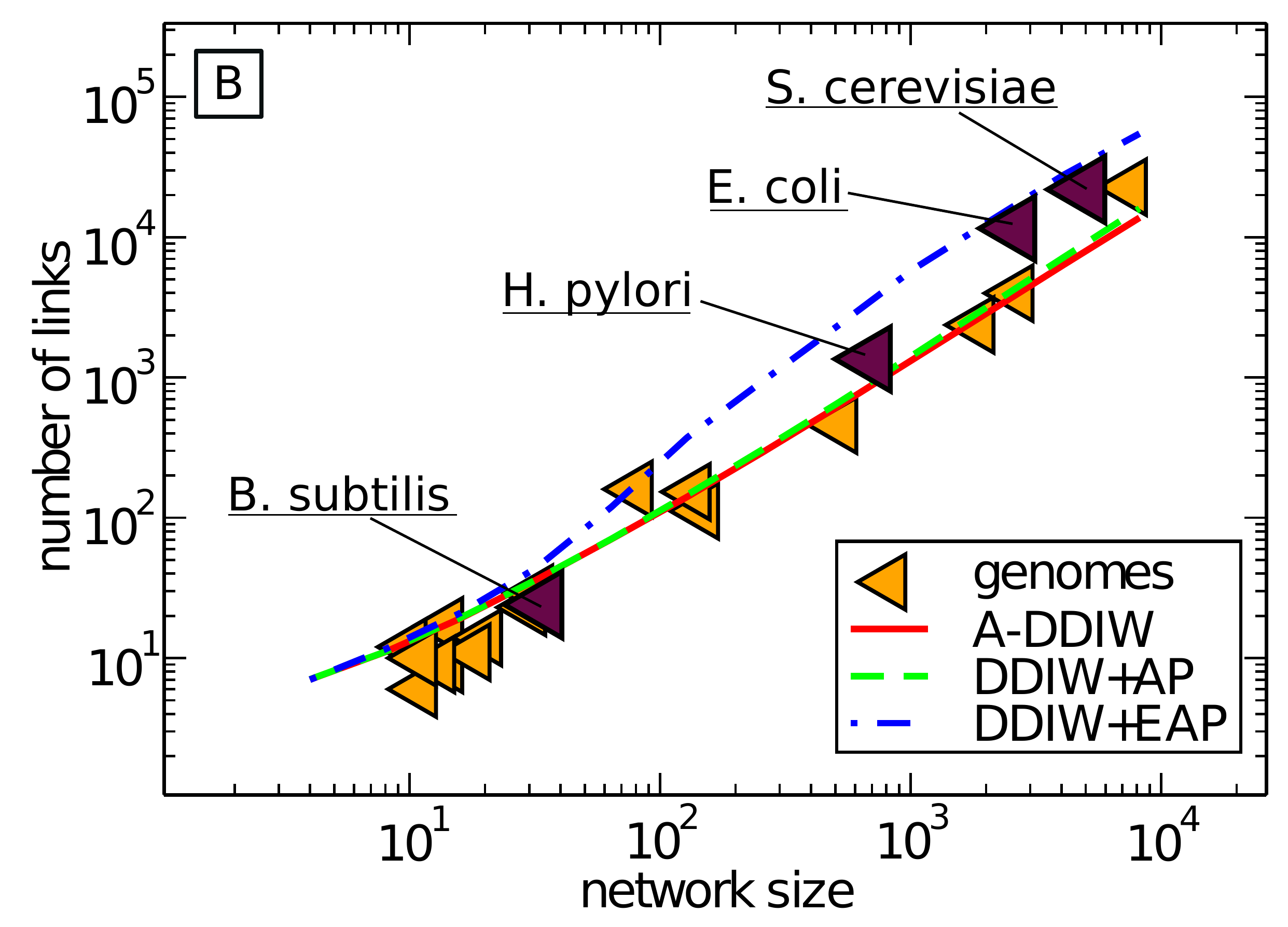}}
    \caption{(Color online) Universal behavior for the number of
      homology classes and for the number of network links.  (A)
      Number of homology classes versus total number of proteins.
      Symbols are data from the SUPERFAMILY database
      \cite{SUPERFAMILY}, the line is a two-parameter $(\alpha,
      \theta)$ fit from the model [equation
        (\ref{eq:crp_asymptotic_theta})].  (B) Number of links versus
      size of sampled network.  Symbols are data from the DIP dataset
      \cite{DIP}, lines are the results of simulations for the
        three variants of the model, with all parameters fixed by
        fits. Darker triangles point out some examples of well-known
      genomes. Note that many networks (e.g. \emph{B.~subtilis}
        and all the triangles of smaller size) are heavily
        undersampled in the data set (see
        Sec.~\ref{section:discussion}).
\label{figure:universal_fits}}
\end{figure}

The calculations presented in the previous section and in the Appendix
are not easily extendable to finite values of $\theta$; they are valid
in the asymptotic limit or when $\theta=0$.  Nonetheless, a corrected
expression of $F(N)$ for the case $\theta>0$ and $F(1)=1$ can be obtained (see
\cite{angelini_amato_etal}), and it is the one we use here to fit
the number of homology classes as a function of genome size,
\begin{equation}
  \label{eq:crp_asymptotic_theta}
  \alpha F(N) + \theta \sim \frac{\alpha+\theta}{(1 +
    \theta^{\alpha})}\left(N+\theta\right)^\alpha .
\end{equation}
We perform the fits on the empirical dataset for homology classes
defined by protein domain architectures, described in
Sec.~\ref{section:data_analysis}.
By taking into account all data, we obtain
$\alpha\simeq0.42$ and $\theta\simeq124$.
Estimates change slightly by imposing a cutoff, since after $N\approx 1000$ data show a clearer power law.
By including only data with $N\geq 1000$, we obtain $\alpha\simeq0.43$
and $\theta\simeq118$, which are compatible with the results obtained
from the whole available range of genome sizes (without any cutoff).
We will use the following estimates for all forthcoming computations:
\begin{equation*}
\alpha=0.43,\quad \theta=121.
\end{equation*}
The theoretical mean-field curve for $F(N)$ is plotted against data in
Fig.~\ref{figure:universal_fits}(A).

Turning to the network data and the fit for $L(N)$, a non-null value of
$\theta$ is not expected to modify the asymptotic behavior, but to act
only on the prefactors.  Therefore we use the mean-field $L(N)$,
even if the homology class fits give a non-negligible value of
$\theta$.
Notice that the same scaling seems to apply to
the prokaryotic genomes as well, despite their network
dynamics not being dominated by duplication divergence; indeed, homology
classes prevalently expand by horizontal gene transfers~\cite{rocha_treangen}.
A more precise analysis of this behavior can only be carried out with more reliable and
abundant data; here we use both prokaryotic and eukaryotic data, as
described in Sec.~\ref{section:data_analysis}.
Fits against the mean-field
predictions for $L(N)$ given in the Appendix (with $\alpha=0.43$
fixed) yield
\begin{equation*}
\sigma=0.457(10)
\end{equation*}
for variant \ref{variant_AP}, and
\begin{equation*}
\begin{aligned}
\sigma=&0.421(9)\quad \left(\gamma=1\right)\\
\sigma=&0.460(10)\quad \left(\gamma\to 0\right)
\end{aligned}
\end{equation*}
for variant \ref{variant_nAP}; values of $\gamma$ between $0$ and
$1$ give estimates between the two extremes.  On the other hand, a fit
against the pure DD prediction (\ref{eq:meanfield_solution_DD}) gives
\begin{equation*}
\sigma=0.446(10).
\end{equation*}
We tested the stability of the foregoing fits by increasing the cutoff on
the network size $N$ from $10$ to $100$. The values do not change appreciably;
errors increase by approximately $50\%$.
Comparison between DIP data \cite{DIP} and simulations of variant
\ref{variant_age}, whose exact behavior cannot be calculated via
mean-field, give approximately $\sigma\simeq0.5$, which corresponds
to an effective link-retention probability around $0.4$
(see Fig.~\ref{figure:meanfield_LN_DDCRP}C).

Figure~\ref{figure:universal_fits}(B) shows numerical results for the
three variants of our model (with $\alpha$, $\theta$, and $\sigma$
fixed by the above fits) superimposed on the data from DIP. The
  initial network was chosen as the complete $3$-graph (see
  Sec.~\ref{section:meanfield} and \ref{section:model_definition}).
Finite-size effects can be seen, especially for the DDIW+EAP variant,
but the trend is consistent. The results for all parameters are
compatible with each other, therefore we regard this as a model
variant-independent fit: the two parameters $\alpha$ and $\sigma$ can
then be seen as ``universal'' (model variant-independent) quantities
governing the scaling laws observed in genomes.  Very similar values
of $\sigma$ ($\approx 0.4$) were also found in
\cite{ispolatov_krapivsky} with a more detailed analysis of the degree
distribution of PPI networks and comparison to the model.  Note that a
simple fit of the form $L(N)\sim N^{2\sigma}$ on the empirical data
would yield $\sigma=0.52$, i.e. it would suggest a crossover regime.
According to our analysis, such a higher exponent appears instead to
be an artifact due to the cooperation of two terms ($N^{2\sigma}$ and
$N$) with smaller exponents but with alternating signs.

Note that in principle the mean-field derivation is valid in the
  large-$N$ limit.
Figure~\ref{figure:universal_fits}(B) shows that differences in the
fit results can be noticed for small cutoffs.
We chose a low cutoff to genomes with less than 10 nodes in order to
show this.  It must be noted however that many ``small'' networks are
actually quite large in reality, but extremely under-sampled in the
data set.

\subsection{Comparison with the empirical network of yeast reveals the
  necessity of age-dependent divergence.}

We now turn to the question of the topological properties and the age
dependency of interactions.  In order to perform a qualitative
comparison between properties of an empirical PPI network and the
results of computer simulations for the three model variants described
above, we choose the case of baker yeast \emph{S.\ cerevisiae}, where
reliable estimates of the age of nodes can be obtained from the
literature (see Sec.~\ref{section:data_analysis}).
As pointed out in \cite{kim_marcotte} and \cite{Liu_Liu_et_al}, while
standard duplication-divergence network growth models well reproduce
topological features of protein-protein interaction graphs, such as
degree distribution and clustering coefficient, they fail to capture
the empirically observed correlation between the evolutionary ages of
interacting proteins. As they discuss, this might be obtained from an
anti-preferential attachment principle, if it becomes a dominant mechanism in
defining the network topology.

In order to monitor topology and history dependency of interactions we
considered the following observables.  (I) We measured two relevant
topology-related quantities: the degree distribution $n_k$, defined as
the fraction of nodes of degree $k$, and a measure of the
degree-degree correlation, called $d_k$, defined as the
average over all nodes of degree $k$ of the mean degree of their
neighbors.  (II) To check for age-age correlations, we employed the
interaction density $D_{m,n}$ and the interaction 
density gradient $\Delta D$ introduced in
Sec.~\ref{section:data_analysis}.

\begin{figure*}[t]
\centering
\includegraphics[scale=.151]{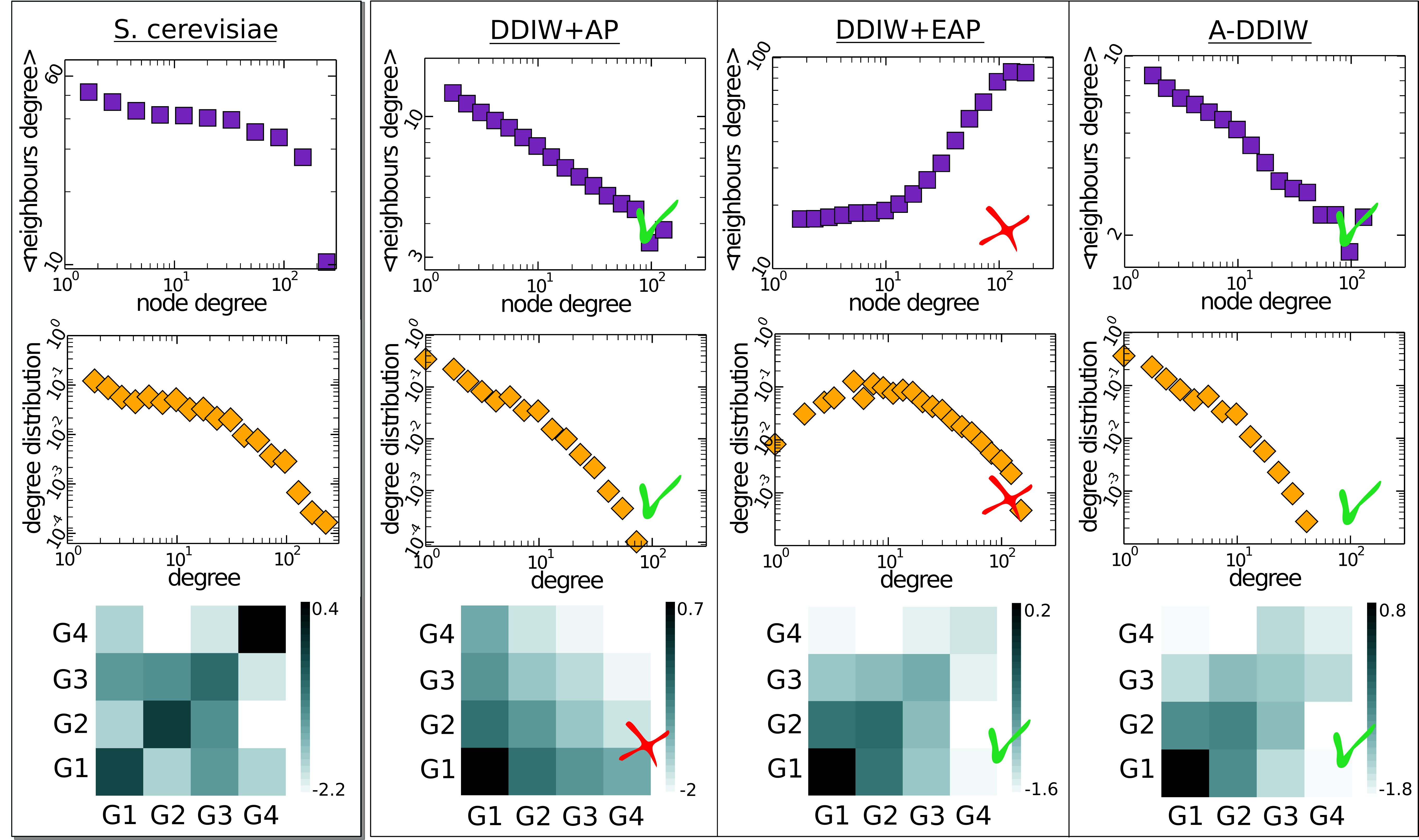}
\caption{
Qualitative comparison between model variants and empirical data.  The
average degree of nearest neighbors (top row), the degree distribution
(middle row), and the interaction density [equation (\ref{eq:interactiondensity})]
between age groups (bottom
row) are measured for \emph{S.\ cerevisiae} (left panel) and for simulations of the
three variants of the DDIW model (right panel).  The DDIW+AP variant
successfully reproduces the empirical degree correlation and degree
distribution, but the wiring mechanism does not
provide enough links to reproduce the empirical age correlation; the
DDIW+EAP variant correctly shows correlation between protein ages,
thanks to the increased number of links introduced by innovation, but
it strongly distorts the topological features of the network; the A-DDIW variant
effectively reproduces both topological and age-correlation features
observed in the empirical network.
\label{figure:comparison}}
\end{figure*}

The behaviour of the observables considered is shown in
Fig.~\ref{figure:comparison} for both the empirical PPI network of
yeast and numerical simulations of the DDIW model variants. The model
parameters are those obtained in Sec.~\ref{section:universal_fits}.
As we pointed out before, results for the age class corresponding to
the WGD in Fig.~\ref{figure:comparison} should be taken carefully,
since homologues in that class were duplicated in a phenomenologically
different event. For assessing how successful each variant is
  in reproducing the degree distribution and the degree correlation we
  adopt a qualitative criterion.  Specifically, we consider a
  monotonically decreasing behavior of the two topological quantities
  to be compatible with empirical data, since this is the behavior
  observed in yeast.  Concerning node-age correlations, we measure the
  interaction density gradient and verify whether it is positive or
  negative; the reference data for yeast give a positive $\Delta D$.
  The whole comparison is carried out in the same spirit as in
  ref.~\cite{kim_marcotte}.

The DDIW+AP variant successfully reproduces the empirical degree
correlation and degree distribution, but not the pattern of
correlation between age groups ($\Delta D<0$).
In this model, the innovation move
gives a negligible contribution to network topology, because the
corresponding number of links is always subdominant. In fact, we verified that
changing the anti-preferential attachment innovation move into
preferential attachment has little or no effect on the main
topological observables. As expected from this argument, this model
generates a network where new nodes are preferentially connected to
old nodes, contrary to the pattern that emerges in yeast, and
equivalently to a pure duplication-divergence network growth.
However, the anti-preferential mechanism is capable of generating a
qualitatively correct age correlation if it can build a large
number of links, i.e.  if the extensive variant DDIW+EAP is considered. In this
case, due to the progressive increasing in the number of links attached in the
innovation move, one obtains the correct empirical age dependency
 ($\Delta D>0$),
but at the expense of completely disrupting the topology. 
For this variant, a scatterplot of the degree-degree correlation (not shown here) 
presents a slight bimodality in a small range of degrees; 
we chose nonetheless to group the data in histograms, in order
to highlight how the overall behavior is different from the empirical one.
Finally, the age-dependent DDIW variant is able to account both for
the topological features and for the age correlation. 

As mentioned above, we have also tested the robustness of the
results under further modifications of the innovation move. No
relevant change in the results for variant \ref{variant_AP} is
detectable by applying a preferential attachment principle instead of
an anti-preferential one, nor by attaching the new node to a fixed
number ($>1$) of existing nodes.  Moreover, variant
\ref{variant_nAP} yields very similar results for all values of
$\gamma$ in $(0,1]$, and therefore the actual value of this parameter
should not be regarded as an essential quantity.  As far as the
age-preference variant \ref{variant_age} is concerned, we remark
that an anti-preferential wiring move gives the clearest results, but
age-age correlations can be seen also in networks obtained by means of
preferential-attachment wiring, as long as this does not dominate over
the duplication-divergence move.

\section{\label{section:discussion} Discussion and Conclusions}

The model presented here can be seen as the prototype of a rather
general modeling framework where a graph grows by the addition of
nodes and links within the constraint of a class structure.  Indeed,
new nodes are added to a new class or to an existing one with
prescribed probabilities, their wiring rules being different in the
two cases.  Here, we explored variants where nodes added together with
a new class are wired to the old network according to an
anti-preferential attachment principle, while nodes introduced into an
existing class follow a duplication-divergence prescription.
The goals of our work were twofold.  First, we studied the
  joint evolution of the network by duplication/divergence and class
  expansion/innovation. Second, as a case study and proof-of-principle
  application, we applied the unified framework to the study of
  age-dependence, where some interesting questions are open. 
  The two objectives are connected, as the scenarios we explored would
  be ill-defined outside of this unified framework.  For example,
  assigning anti-preferential attachment to the innovation move
  requires to be able to distinguish it from a duplication move,
  i.e. to separate new families from existing ones. To carry out both
  objectives, we stayed as close as possible to empirical data.

We considered probabilities of addition of new nodes that vanish with
$N \to \infty$, in order to reproduce the observed empirical scaling
of homology classes \cite{cosentinolagomarsino_sellerio_etal_2009}. As
a consequence, unless it is imposed that new nodes (i.e. new nodes
belonging to a new homology class) carry an extensive number of links,
the wiring rule for innovation is of secondary importance with respect
to the duplication-divergence move in determining the asymptotic
features of the resulting graph ensemble. This is in accordance with
the empirical observations indicating that duplication-divergence is
relevant in shaping the appearance of the PPI
network~\cite{Pereira-Leal2005,Presser2008,Teichmann2002,Wang2011}.
The finite-size behavior, nonetheless, is sensitive to the
  innovation process, suggesting the existence of non-trivial features
  of the topology related to the dynamics of homology classes.

\begin{figure}[t]
\centering
\includegraphics[scale=0.35]{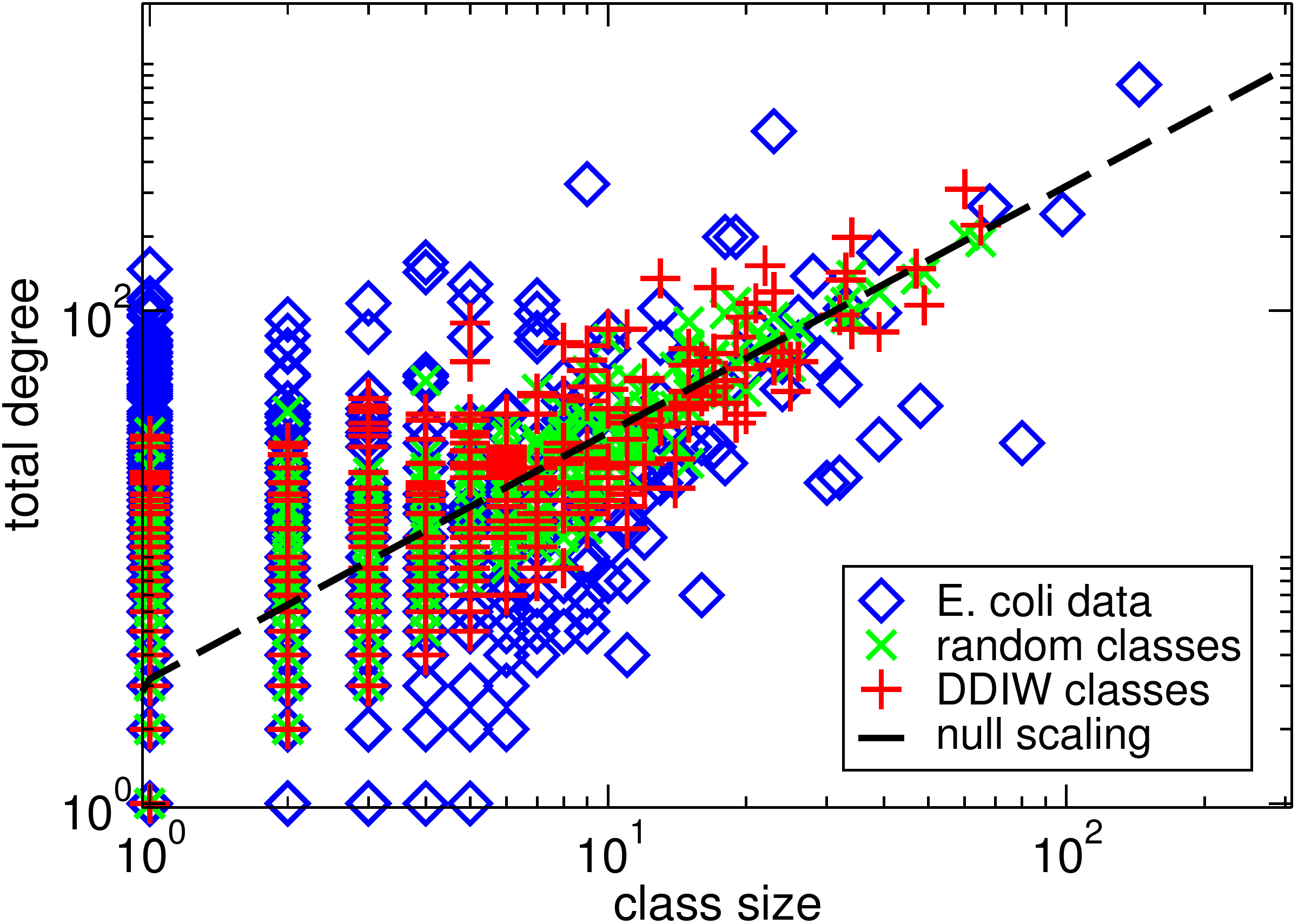}
\caption{(Color online) Linear scaling of the correlation between
the total number of links originating from a given
class and the size of the class:
scatterplot of class degree (sum of the degrees of all nodes in a class)
versus class size (number of nodes).
Red crosses ($+$) represent results from the typical realization of the DDIW model
with $N=2640$ nodes as in \emph{E.\ coli}'s PPI network
(for the DDIW+AP variant; A-DDIW yields a completely similar plot);
green crosses ($\times$) are obtained from the same DDIW realization by randomly permuting
nodes between classes; blue diamonds are obtained by combining data for
network structure and homology classes for \emph{E.\ coli};
the dashed line is the prediction for the average of the total class degree
in the randomized case, \ie the mean node degree times the class size (here $\left<k\right> = 2.3$).
\label{figure:joint}}
\end{figure}

Following these indications, the framework considered here can
  in principle make more detailed predictions for observables that
  involve network and homology classes jointly.  We analyzed the
  behavior of one such observable, namely the correlation between the
  total number of links originating from a given class and the size of
  the class.  While we find good agreement between data for the
  \emph{E.\ coli} PPI network and simulations of the DDIW model (at
  least for the two non-extensive variants), they both agree with the
  null expectation that this scaling is linear (see
  Fig.~\ref{figure:joint}).  Indeed, in the random case, i.e. when the
  members of homology classes (of prescribed sizes) are chosen
  randomly among network nodes, the total degree of a class will be,
  on average, equal to the number of nodes in the class times the mean
  node degree. Thus, we were unable to find such an effect in the data
  available to us.

Despite the relation between class size and total degree not being
discriminating, the DDIW model does generate nontrivial correlations
from the joint evolution of the network and the partitioning into
classes.
The fact that currently available empirical data do not allow to
discriminate should not, in our opinion, discourage the analysis of
joint models until more abundant or precise data will be
available.
To give an example, let us focus on the number $F_N(1,1)$ of classes
containing a single node with degree $1$ in a network of size $N$.  In
the null model where nodes are shuffled randomly between classes (in a
single realization of the network), this number is distributed
following a hypergeometric distribution ``centered'' in
$\left<F_N(1,1)\right>=MC/N$, where $M$ is the number of degree-one
nodes in the whole network, and $C$ is the number of size-one classes.
Simulations of the DDIW model for several realizations (in the two
non-extensive variants) consistently yield values of $F_N(1,1)$ that
lie several standard deviations above the mean of the null-model
distribution.  We have measured $M,C,N$, and $j$ for the \emph{E.\ coli} PPI
network, both using Ensembl \cite{ensembl} and SUPERFAMILY
\cite{SUPERFAMILY} homology data; the actual value of $F_N(1,1)$ is
larger than the null average, in both datasets, by approximately
$4\sim6$ standard deviations, thus confirming the qualitative non-null
prediction of the model.  Future work could be directed towards a more
detailed study of joint laws such as this one.  As an example, the
full numbers $F_N(i, k)$ of classes containing $i$ nodes of total
degree $k$ are a class of interesting observables which are probably
accessible by standard mean-field techniques.

The model variants can be approached by analytical estimates and
direct simulation, and matched with empirical data on both homology
classes and PPI networks. This fitting procedure constitutes a proof
of principle of the general applicability of the framework defined
here. It also allows to fix the few parameters of the model, and
produces well-defined comparisons of the model's predictions with
data.

In order to explicitly carry out this comparison in a specific case
study, we considered the problem of reproducing the empirical age
dependency of PPI network interactions through different variants of
the model. We tested the predictions obtained against data from yeast,
where both PPI network and gene duplications are well characterized,
and the duplication age of individual proteins is also available.
We were able to show that the empirical duplication age
patterns of interacting protein pairs can be reproduced in two
alternative ways. First, by an anti-preferential attachment
prescription in the innovation move, associated to a heavy (extensive)
contribution of this move to the number of links. Second, by inserting
a strong negative bias towards forming protein-protein interactions
with old nodes. However, the first choice leads to networks whose
degree distribution and neighbour degree correlations do not resemble
the empirical ones.  Conversely, the bias imposed in the second case
could be rationalized by biological arguments concerning the available
binding interfaces (older proteins are more likely fully engaged with
the interactions they participate into) and the conservation of basic
biological functions (new interactions interfering with older ones
could be detrimental). Thus, an age-dependent duplication-divergence
move seems more satisfactory.
Once established that such an age dependence in the divergence
  process is in qualitative agreement with data, one can ask whether
  the same features can be reproduced without considering the full
  partition/topology dynamics.  We have performed additional numerical
  simulations and found that the qualitative patterns in
  Fig.~\ref{figure:comparison} can be reproduced also by a simple
  duplication-divergence model with age bias and no innovation nor
  class dynamics.  This is not in contrast with the importance of
  considering the problem in the more general framework, since in
  principle --- as we have explained in the previous section --- other
  mechanisms, related to the innovation/wiring move, could have been
  responsible for the age correlation patterns observed.

Overall, our analysis tends to support the hypothesis that
duplication-divergence alone does not account for the observed
history-dependency of the existing protein-protein
interactions~\cite{kim_marcotte}.  Note, however, that in the
age-dependent DDIW model, as well as in the previous models of this
kind, duplication-divergence turns out to be a necessary ingredient in
shaping biologically-resembling degree distributions and degree
correlations of nearest neighbors. This suggests that the
  mechanism of duplication and divergence might play a role in
  determining PPI network topologies~\cite{Navlakha2011}.
  Conversely, in the previous model of Kim and Marcotte, the age
  dependency is associated to model moves that, roughly speaking, are
  more similar to an anti-preferential attachment innovation move than
  to a duplication-divergence one~\cite{kim_marcotte}.
  We should also remark that the models we have explored here are
  based on totally asymmetric duplication-divergence.  We cannot
  exclude that the age-correlation patterns could be biased also by
  using general duplication-divergence schemes
  \cite{evlampiev_isambert}, where different values of sigma are
  assigned to the connections between pairs of new nodes with respect
  to new-old node pairs. In this case, the introduction of an
  additional parameter could produce the age correlation kernel in a
  natural way.

One important caveat is that the PPI data available to us are affected
by strong sub-sampling problems, since presumably for most organisms
only a fraction of the protein-protein interactions are
available in the DIP database~\cite{DIP}.  Having small samples of
large networks makes it problematic to estimate model parameters. For
example, it is likely that the exponent for $L(N)$ is overestimated.
We performed a numerical test by growing networks up to size $N$ (and
a fluctuating number of links $L'$) and sumbsampling them to a fixed
number of links $L$. In general one obtains networks with many more
nodes ($N'$) compared to networks that are grown with the model at
$L'$ edges and not subsampled.  For parameter values that match the
available data, this error could be as large as $100\%$; in C.\
elegans, for instance, for which approximately $4000$ interactions are
known involving around $2600$ proteins (out of $\approx 20000$ genes),
we obtain $N'\approx 5100$.  
On the positive side, restricting the parameter-matching
  analysis (Figure~\ref{figure:universal_fits}) of the model to the
  few highly sampled genomes does not change our results.
Nevertheless, it seems quite possible that a larger
cross-genomic knowledge of PPI networks could change the quantitative
picture emerging from these data, and possibly also the
qualitative one.

To conclude, despite of the current open questions, we believe that
this general framework might be important to pose questions about the
growth of PPI networks, as the network structure is intimately related
to the partitioning in homology classes, and, quite importantly, to
the class of biological functions that a specific homology class can
perform~\cite{grilli2012}.\\

\acknowledgments{ We thank Alessandro Sellerio for help with the
  construction of homology classes from protein domain architectures.
  We are grateful to Herv\'e Isambert and Sergei Maslov for
    useful comments and suggestions.  M.G. acknowledges financial
  support from Fondo Sociale Europeo (Regione Lombardia), through the
  grant ``Dote ricerca''.  }

\appendix*

\section{Mean-field calculation of $L(N)$}
We give here the solutions to the mean-field equation (\ref{eq:ODE_DDCRP}).
Let us call $L_\mathrm{\ref{variant_AP}}(N)$ the solution with the choice $l(N)=1$
(variant \ref{variant_AP}) and $L_\mathrm{\ref{variant_nAP}}(N)$ the solution 
with the choice $l(N)=\gamma 2 L/N$ (variant \ref{variant_nAP}).
For both choices (\ref{eq:ODE_DDCRP}) is a standard
first-order ordinary differential equation, whose solution can be readily
computed with the help of \emph{Mathematica}.
One obtains
\begin{equation}
\label{eq:L_A_fullform}
\begin{aligned}
L_\mathrm{\ref{variant_AP}}(N)&=N^{2\sigma}\rme^{2\sigma P_\alpha(N)}\Bigg\{
\mathrm{const.}+\frac{1}{2}N^{1-2\sigma}\rme^{-2\sigma P_\alpha(N)}\\
&\phantom{=}-\frac{1}{2(1-\alpha)}N^{1-2\sigma}
\left[2\sigma P_\alpha(N)\right]^\frac{1-2\sigma}{1-\alpha}\times\\
&\phantom{=}\times\Gamma\left(-\frac{1-2\sigma}{1-\alpha},2\sigma P_\alpha(N)\right)\Bigg\},
\end{aligned}
\end{equation}
where $P_\alpha(N)$ is defined as
\begin{equation}
P_\alpha(N)=\frac{\alpha}{1-\alpha}N^{\alpha-1}
\end{equation}
[which is proportional to the asymptotic form of the innovation probability,
see Eq.~(\ref{eq:asymptotic_probcrp})], and $\Gamma(a,z)$ is the upper incomplete gamma function
\begin{equation}
\Gamma(a,z)=\int_z^\infty t^{a-1}\rme^t \,\rmd t.
\end{equation}
The constant term depends only on $\alpha$, $\sigma$, and the initial condition $L(N_0)=L_0$.
Notice that $P_\alpha(N)\to 0$ when $N\to\infty$, since $\alpha\in(0,1)$.
By substituting the asymptotic expansion for the incomplete gamma to leading order around $z=0$
\begin{equation}
\Gamma(a,z)\sim\Gamma(a)-\frac{z^a}{a}
\end{equation}
into (\ref{eq:L_A_fullform}) one sees that the first term in curly brackets gives a contribution $\propto N^{2\sigma}$
to the asymptotic form, while the second and third terms have a linear behavior $\propto N$, thus
recovering expression (\ref{eq:asymptotic_LN_AP}).

A similar, but more complicated, expression to equation
(\ref{eq:L_A_fullform}) is found for
$L_\mathrm{\ref{variant_nAP}}(N)$; we do not quote it
  here because it is very large without being particularly
  instructive; the same analysis gives the corresponding asymptotic
behavior (\ref{eq:asymptotic_LN_nAP}).  \vfill

\begin{table}
\begin{tabular}{|l|r|r|}
\hline
Organism & \# Nodes & \# Links \\
\hline
Sulfolobus solfataricus &       14 &        9 \\
\hline
Arabidopsis thaliana &      136 &      153 \\
Bos taurus &       30 &       23 \\
Caenorhabditis elegans &     2647 &     3985 \\
Chlamydomonas reinhardtii &       14 &       17 \\
Danio rerio &       13 &        9 \\
Drosophila melanogaster &     7500 &    22737 \\
Gallus gallus &       11 &        6 \\
Homo sapiens &     1850 &     2370 \\
Mus musculus &      524 &      457 \\
Pisum sativum &       10 &       12 \\
Rattus norvegicus &      147 &      112 \\
Saccharomyces cerevisiae &     4998 &    21881 \\
Schizosaccharomyces pombe &       80 &      160 \\
Xenopus laevis &       20 &       14 \\
\hline
Bacillus subtilis &       34 &       24 \\
Caulobacter crescentus &       18 &       11 \\
Escherichia coli &     2640 &    11545 \\
Helicobacter pylori &      700 &     1354 \\
Mycobacterium tuberculosis &       13 &        9 \\
Synechocystis sp. &       32 &       29 \\
Xanthomonas campestris &       11 &       10 \\
\hline
\end{tabular}
\caption{
  Genomes from DIP \cite{DIP} and corresponding values of 
  the number of nodes $N$ and number of links $L$.
  \emph{Archaea} on top, then \emph{prokaryotes}, then \emph{bacteria}
  (separated by horizontal lines). Note that the bacteria with
    small number of nodes are heavily undersampled in the data set, so
    that the number of effectively significant points is low (see
    Sec.~\ref{section:discussion}).
\label{table:genomes}
}

\end{table}

\bibliography{bibpaper}

\end{document}